# A NOVEL ALGORITHM FOR THE ESTIMATION OF THE SURFACTANT SURFACE EXCESS AT EMULSION INTERFACES

German Urbina-Villalba[1]*, Sabrina Di Scipio[2], Neyda García-Valera[1]

[1]Laboratorio de Fisicoquímica de Coloides. Centro de Estudios Interdisciplinarios de la Física (CEIF) del Instituto Venezolano de Investigaciones Científicas (IVIC). Aptdo. 20632, Email: guv@ivic.gob.ve
[2]Departamento de Termodinámica y Fenómenos de Transferencia. Universidad Simón Bolívar.

**Abstract**      When the theoretical values of the interfacial tension -resulting from the homogeneous distribution of ionic surfactant molecules amongst the interface of emulsion drops- are plotted against the total surfactant concentration, they produce a curve comparable to the Gibbs adsorption isotherm. However, the actual isotherm takes into account the solubility of the surfactant in the aqueous bulk phase. Hence, assuming that the total surfactant population is only distributed among the available oil/water interfaces, one can calculate what surface concentration is necessary to reproduce the experimental values of the interfacial tension. A similar procedure can be followed using the zeta potential of the drops as a standard for a given set of salt and surfactant concentrations. We applied these procedures to the case of hexadecane/water nanoemulsions at different salt concentrations. This information was used to compute typical interaction potentials between non-deformable nanoemulsion drops. The results indicate that there are significant differences between the surfactant population expected from macroscopic adsorption isotherms, and the actual surfactant population adsorbed to the surface of nanoemulsion drops.

## 1. INTRODUCTION

Recent experimental evidence indicates that the interfacial area occupied by a surfactant at the surface of a nanoemulsion drop might be one order of magnitude higher than the one usually exhibited by the same molecule at macroscopic interfaces [De Aguiar, 2010; De Aguiar, 2011]. The measures spanned 1% v/v oil-in-water nanoemulsions of hexane, dodecane and hexadecane with a mean radius between 80 and 130 nm. The origin of such difference has been initially adjudicated to the lack of equilibrium between the surrounding liquid solution and the interface of the drops.

The implication of this inconsistency is relevant, since it is the amount of surfactant adsorbed which determines the repulsive interaction between the drops, and hence, their time scale for aggregation. In fact, it is both surprising and perturbing, that a great deal of effort and financial support is focused on the synthesis of nanoemulsions for biomedical applications, while little effort is dedicated to understand the origin of their instability. The current know-how of nanoemulsion stability is limited to the implementation of elementary concepts of DLVO theory [Verwey, 1948; Derjaguin, 1941] for flocculation control, and the addition of insoluble substances to the internal phase of the emulsion, in order to regulate the Ostwald ripening effect [Kabalnov, 2001].

Oil-in-Water (O/W) nano-emulsions as any other emulsions are systems out-of-equilibrium whose stability has a kinetic origin. They separate completely into their immiscible phases through several mechanisms, whose relative importance is mainly determined by the size of the drops, the solubility of the internal phase, the surfactant concentration and the type of surfactant molecules employed.

Recently we developed [Rahn-Chique, 2012] an analytical procedure which assumes a constant kernel of aggregation rates [Smoluchowski, 1917] to evaluate the average flocculation rate of nanoemulsions. Both the experimental procedure and the theoretical approach are robust, and their extension for the assessment of Ostwald ripening from a non-classical point of view appears feasible. In any event, and despite the utility of these approaches they are phenomenological in nature, and therefore, they cannot decipher the origin of the observed behavior. The reason for this limitation is linked to the fact that all destabilization processes of liquid dispersions are surfactant-dependent, and the distribution of surfactants between the interface of the drops and the bulk phases is continually evolving.

Going back to our original statement regarding the actual surfactant concentration at a nanoemulsion interface, one may wonder if the amount of interfacial area generated during the synthesis of nanoemulsions, can affect significantly its adsorption equilibrium. Our general knowledge of adsorption isotherms is built upon our observations of macroscopic interfaces. Under these conditions it was established –for example-, that surfactant molecules generally do not aggregate before the oil/water interface is completely saturated. This occurs at a specific surfactant concentration known as the critical micelle concentration (CMC). However, if the very same system is vigorously stirred in order to make a nanoemulsion, the increase of interfacial area is considerable, and it is unclear if the same surfactant concentration is still capable of producing a similar surface excess ($\Gamma$) at the interface of the drops. It appears that a much higher surfactant concentration is now required for this purpose, and hence, it is likely that the effective interfacial area per surfactant molecule is considerably higher, as recently suggested by the group of Roke [De Aguiar, 2011].

## 2. THEORETICAL BACKGROUND

### 2.1. Major difficulties in the simulation of the stability of liquid/liquid dispersions:
1. The existence of a time-dependent interaction potential
2. The impossibility to move surfactants and drops in the same time scale

In Emulsion Stability Simulations (ESS) the movement of the oil drops is mimicked using a simplified version of the Brownian Dynamics (BD) algorithm from Ermak and McCammon [Ermak, 1978; Urbina-Villalba, 2000; Urbina-Villalba, 2004a; Urbina-Villalba, 2009]. In our simulations the suspending liquid is not explicitly considered, but its effect on the movement of the particles is incorporated by means of two terms: 1) a deterministic force which is a scaled by the "effective" diffusion constant of the drops in the aqueous media, and 2) a stochastic force with certain statistical properties which summarizes the effect of the millions of collisions of the water molecules on the surface of the drops. These two contributions are linked through the fluctuation-dissipation theorem. The theorem connects the effective diffusion constant of the drops to the random displacement of the particles, in such a way that the average thermal energy of the system is constant and of the order of $k_BT$, where $k_B$ is the Boltzmann constant, and T the absolute temperature.

It was early demonstrated by Dickinson and co-workers [Dickinson, 1979; Bacon, 1983] that the formal computation of hydrodynamic interactions using pair wise contributions was both inaccurate and time consuming. Moreover, it was also evident that the potential of interaction between the drops was usually shorter than 100 nm, which requires very short times steps in order to sample the relevant features of the interaction potential appropriately. This is especially important if one deals with liquid drops. Coalescence is an irreversible process in quiescent media. Since aggregation is a necessary step in the route to coalescence, the outcome of a simulation might drastically change if the repulsive potential between the drops is not observed due to an inconvenient selection of the time step ($\Delta t$). In fact, a large displacement (long $\Delta t$) is equivalent to a "jump" over the repulsive barrier between two drops, since the repulsive potential has no influence in the calculation of the force neither before nor after the particle movement.

Typically, a suitable time step is of the order of $10^{-7}$ seconds, which means, that one second of actual time requires 10 millions iterations of the program. It is also important to remark that the standard algorithm of Brownian Dynamics developed by Ermak and McCammon [Ermak, 1978] poses an additional restriction on the minimum time step allowed. This limit is necessary in order to guarantee equilibrium in the momentum distribution of the drops. Such restriction is inconvenient and attempts towards the correct sampling of the interaction potential. Moreover, it is pointless in the case that other destabilization processes –like coalescence- are considered, because in this case no account of the momentum conservation is carried on during the process.

In the case of BD as in all Molecular Dynamics (MD) calculations, the repulsive potential between the particles is fixed. It results from the chemical composition of their surfaces and the pH of the suspending media. The particles are invariably rigid. The movement of the particles is calculated assuming pair wise interactions, and obtaining the deterministic forces from the differentiation of the potential as a function of the interparticle distance. In the case of DLVO, the interaction potential involves two contributions: a) an attractive potential due to the dipolar interaction between the molecules of the particles, and b) a repulsive potential due to the presence of electrostatic charges at their surfaces. These potentials generate a repulsive barrier towards aggregation with one minimum on each side. Hence, the outcome of a simulation can only be a particular degree of flocculation of the particles depending on the physicochemical conditions. In particular, the salt concentration regulates the height of the maximum and the depth of the secondary minimum.

In emulsions, the surface charge of the drops depends on the degree of surfactant adsorption. However, the size of the drops changes as a function of time due to coalescence and Ostwald ripening. Consequently, the total interfacial area of the dispersion is continuously modified. If the interfacial area increases, more surfactant molecules are needed to cover the interface. If it decreases, the bulk concentration of these molecules increases, either directly through the detachment of molecules from the interface, or indirectly through the relaxation of a non-equilibrium surfactant excess at the interface of the remaining drops. In general, it is expected that any non-homogeneous surfactant distribution -produced by any arbitrary destabilization mechanism- should rapidly relax. The new value of $\Gamma$ will depend on the total interfacial area of the emulsion and the total surfactant concentration. These changes in the adsorption equilibrium of the surfactant cause changes in the repulsive potential between the drops. The nature of this repulsive force depending on the chemical structure of the surfactant adsorbed. In the case of ionic surfactants, each molecule adsorbed adds its charge to the interface of the drop, generating its repulsive electrostatic potential. Hence, it is necessary to account for the surface concentration of surfactants in the drops at each step of the simulation.

Further considerations regard the attainment of the adsorption equilibrium itself. De Aguiar et al. [De Aguiar, 2011] suggested that the significant movements made by a nanodrop during a millisecond, works against the accomplishment of an adsorption equilibrium similar to that of macroscopic interfaces which requires at least several seconds. Therefore, even in the absence of other destabilization process it is uncertain if the surface excess of the surfactant at the interface of the drops should correspond to the one predicted by the adsorption isotherms.

Hence, it is the physical differences between suspensions and emulsions, which lead to the most important distinctions between their simulation techniques. ESS concentrates on the processes of flocculation, coalescence, creaming and Ostwald ripening. The interaction potential between the drops results from the adsorption of surfactant molecules to their interfaces. This adsorption essentially depends on the total surfactant concentration, the salinity of the aqueous medium, the temperature, the total amount of interfacial area of the emulsion, and *time*. As previously discussed, this time-dependence is not necessarily related to the process of adsorption itself –which requires a finite time [Bonfillon, 1994; Urbina-Villalba, 2004b]-. It is rather connected to the redistribution of surfactant molecules throughout the system, caused by the effect of the mechanisms of destabilization of the emulsion.

The major obstacle in the consideration of the surfactant adsorption in simulations is the fact that it is not possible to move the surfactant molecules and the drops using the same time step. This is the main reason why the molecular description employed by Computational Chemistry programs can only calculate the properties of the interface, but cannot simulate their effect on the stability of oil/water dispersions [Urbina-Villalba, 1995; Urbina-Villalba, 1996]. The rotation of an alkyl chain is of the order of picoseconds ($10^{-12}$ s), and ESS are already extremely slow considering the movement of drops alone at $\Delta t = 10^{-7}$ s. Hence, a molecular description of the whole system is discarded. Moreover, the typical diffusion constant of a molecule in a liquid is of the order of $10^{-10}$ m$^2$/s, while the one of a nanodrop is of the order of $10^{-12}$ m$^2$/s ($10^{-14}$ m$^2$/s for a macro emulsion drop). Therefore, a mesoscopic description of the drop movement is too slow in comparison to the diffusion of surfactant molecules. A possible solution to this dilemma is to simulate the effects caused by the diffusion of surfactant molecules rather than to simulate their movement. In regard to emulsion stability, it is the change of the interfacial properties of the drops which is important. Hence, if the interfacial properties can be changed in a way consistent with the experiment, ESS would retain its predictive power, without moving the surfactant molecules explicitly. This modus operandi preserves the spirit of the original Brownian Dynamics algorithm in which the solvent in considered implicitly through the effective diffusion constant of the drops.

## 3. COMPUTATIONAL DETAILS

### 3.1 Apportioning the surfactant population among the drops in emulsion stability simulations. Case 1: a distribution strategy for the equilibrium distribution of an ionic surfactant in the absence of salt.

The procedure described in the last paragraph of the previous section was implemented in ESS through a set of routines that "distribute" the surfactant molecules among the available interfaces using different approximations, each of which is meant to resemble a typical experimental situation (inhomogeneous surfactant distributions [Urbina-Villalba, 2001b], time-dependent surfactant adsorption [Urbina-Villalba, 2001a; Urbina-Villalba, 2004b], etc).

Let us illustrate the procedure outlined using the simplest distribution strategy: the one that assumes a homogeneous and instantaneous distribution of the surfactant molecules amongst the interface of the drops up to a saturation limit ($A_{s,min} = 50$ Å$^2$ [Rosen, 1989]). The interfacial area considered for this specific example corresponds to the one resulting from a Gaussian distribution of drops with an average radius of 184 nm ($\phi = 9.5$ x $10^{-4}$, $n_0 = 3.63$ x $10^{16}$ drops/m$^3$). The program calculates the surface excess of the drops (number of molecules per unit area) using the following scheme:

1) It computes the total interfacial area ($A_T$) of the emulsion, and calculates the total amount of surfactant required ($N_{req}$) in order to cover all the drops up to a saturation limit:

$$A_T = \sum A_i = 4\pi \sum R_i^2 \qquad (1)$$

$$N_{req} = A_T / A_{s,min} = \sum \left( A_i / A_{s,min} \right) = \sum \left( N_{max,i} \right) \qquad (2)$$

2) It calculates the available surfactant population from the total surfactant concentration of the system and the volume of the simulation box:

$$N_T = C_s V_T \qquad (3)$$

3) If there are enough surfactant molecules to cover completely the interfacial area of the drops ($N_T \geq N_{req}$), it assigns the maximum surfactant concentration to each drop:

$$N_{s,i} = A_i / A_{s,min} = N_{max,i} \qquad (4)$$

4) If on the contrary $N_T < N_{req}$, it apportions the total surfactant population according to the relative interfacial area of each drop:

$$N_{s,i} = N_T (A_i / A_T) \qquad (5)$$

5) Finally, the interfacial properties including, the tension ($\gamma_i$) and the charge ($Q_i$) of each drop are computed from the amount of surfactant adsorbed:

$$\gamma_i = \gamma_0 + (\gamma_{CMC} - \gamma_0)(N_{s,i} / N_{max,i}) \qquad (6)$$

$$Q_i = -q_i \, e N_{s,i} \qquad (7)$$

where $\gamma_0$ and $\gamma_{CMC}$ correspond to the value of the interfacial tension in the absence of surfactants, and at $C_s$ = CMC, and $e$ is the unit of electrostatic charge (1.6 x $10^{-19}$ Coul).

In a few words, this simple routine distributes the total amount of surfactant molecules between the available interfaces of the drops up to a saturation limit. Hence it does not consider the amount of surfactant which needs to be dissolved in the bulk phase in order to establish the adsorption equilibrium of the surfactant between the external phase of the emulsion and the interface. However, the algorithm described above takes into account how a change in the total interfacial area of the emulsion changes the surfactant population of the drops: if the number of drops decreases, $A_T$ decreases, and algorithm promotes a slight increase in the charge of the remaining drops. This increase only occurs whenever the initial surfactant concentration is not enough to cover the drops up to their saturation limit, and it is an indirect consequence of the reduction of the total interfacial area of the emulsion.

The adsorption isotherm resulting from the implementation of the routine described above is represented by the left-hand-side curve of Figure 1. We will refer to this theoretical model of adsorption, as Model 1. Notice that the algorithm is sufficient to calculate the repulsive electrostatic potential energy (and the force) between two drops, if the value of $q_i$ (Eq. 7) is known. The electrostatic energy ($V_E$) depends on the electrostatic potential of each drop $V_E = V_E(\Psi_1, \Psi_2)$, which can be calculated given the charge of the drop $\Psi_i = \Psi_i(Q_i)$ [Urbina-Villalba, 2006]. Since Eqs. (5) and (4) provide the value of $N_{s,i}$, the value of $q_i$ can be obtained using an experimental measurement $\Psi_i$ to relate the electrostatic potential to the total charge of the drop. In general $\Psi_i$ is approximated by the value of the zeta potential of the drop obtained from electrophoresis.

The curve on the right-hand-side of Figure 1 corresponds to the experimental results of Rehfeld [Rehfeld, 1967] for the adsorption of sodium dodecyl sulfate at a macroscopic heptadecane/water interface. As shown by this author, the tension of this system can be expressed as an analytical function of the natural logarithm of the total surfactant concentration:

$$\gamma = -1.62(\ln C_s)^2 - 31.5 (\ln C_s) - 102.3 \qquad (\text{mN/m}) \qquad (8)$$

It is remarkable that the general shape of the theoretical curve resembles the one of the actual isotherm. The fact that the curve predicted by Model 1 is moved toward lower surfactant concentrations is a consequence of apportioning the total surfactant population among the interfaces of the drops exclusively.

The similarities between the actual isotherm and the one predicted by Model 1, led us to wonder if it was possible to reproduce the exact experimental data of Rehfeld using the homogeneous surfactant distribution algorithm. For this purpose it was only necessary to change the total surfactant concentration systematically until the experimental values of Rehfeld were recovered. If the method works, this surfactant concentration would be equal to the amount of surfactant adsorbed to the interface in the actual system. The results of this procedure correspond to the "square" symbols that lie on top of the right-hand-side curve in Figure 1. As expected, the data of Rehfeld is completely recovered. Consequently, the interfacial concentration of surfactant necessary to reproduce the experimental adsorption isotherm is obtained. The bulk phase concentration of SDS "at equilibrium" corresponding to each value of the interfacial tension being equal to the difference between the total surfactant concentration suggested by Rehfeld, and the surfactant concentration predicted by Model 1.

Figure 2 shows the surface surfactant concentration obtained, as a function of the total surfactant concentration ($C_s$). The curve can be adjusted to a polynomial of fourth degree with a regression coefficient of 0.9992. It is remarkable that the curve increases monotonically resembling the shape of a Langmuir adsorption isotherm.

Figure 3 illustrates the variation of the bulk surfactant concentration at equilibrium ($C_{eq}^{bulk}$). The surface concentration is so small with respect to the overall surfactant population, that the bulk concentration comes out to be linear, with a regression coefficient of 0.999998. This result is consistent with previous studies of Aveyard et al. [1986; 1987; 1988] regarding the distribution of sodium diethyl-hexyl sulphosuccinate (AOT) between heptane and water. In these measurements, heptane and aqueous solutions of NaCl were mixed together with the surfactant molecules initially dissolved in either phase. The resulting emulsion was left to separate for a few days in a thermostat, until two immiscible phases were observed. The surfactant concentrations were determined by titration of the aqueous phase with hyamine [Aveyard, 1986]. The surface excess was not quantified and assumed to be negligible in comparison to the bulk concentration. The results indicated that for salt concentrations of 0.0171, 0.0513 and 0.1027 M, all the surfactant concentration resided in the aqueous phase up to the CMC. Thus, a plot of $C_{eq}^{bulk}$ vs. $C_s$ showed a straight line of unit slope analogous to Figure 3. A similar situation was observed above the CMC for low salt concentrations (unable to generate a third surfactant-rich middle phase). Otherwise, the surfactant concentration in aqueous phase remained constant above the CMC, meaning that the excess surfactant molecules were been transferred to the middle phase.

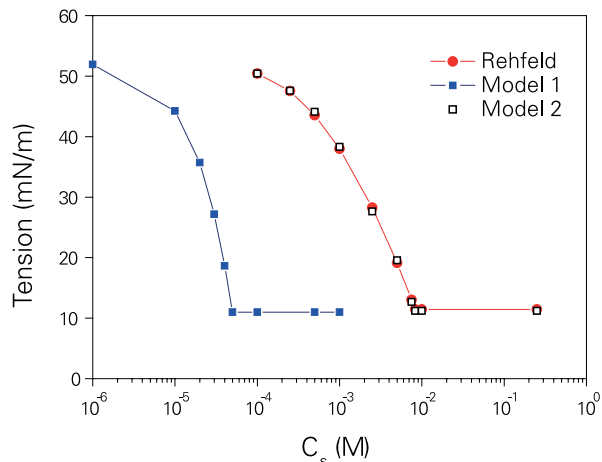

Figure 1: Generic adsorption isotherms of sodium dodecyl sulfate at an oil/water interface.

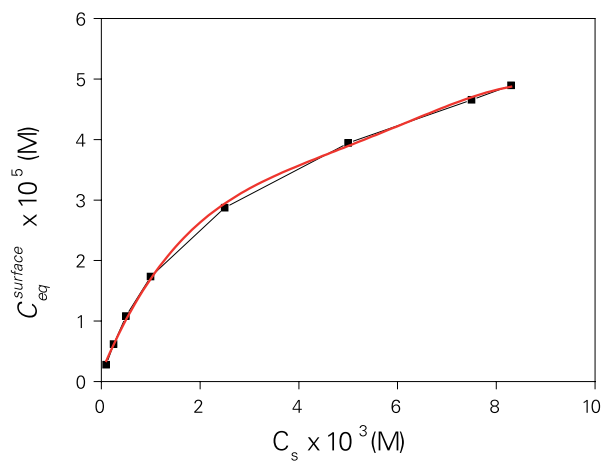

Figure 2: SDS surface concentration predicted by Model 2 as a function of the total surfactant concentration in the system.

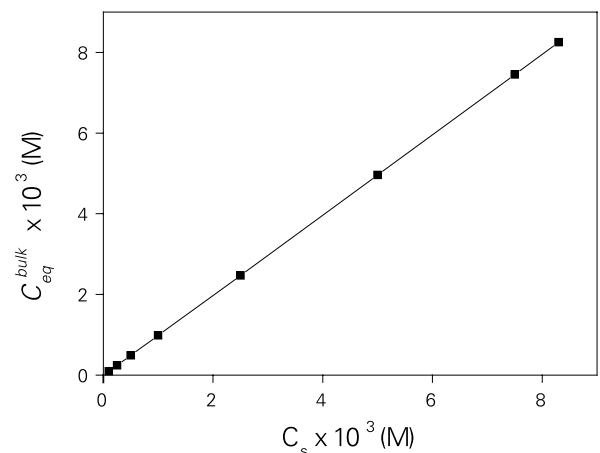

Figure 3: SDS bulk concentration predicted by Model 2 as a function of the total surfactant concentration in the system.

Probably the most remarkable feature of the procedure outlined above regarding the computation of the surfactant surface excess using Model 1 and an experimental adsorption isotherm, is that it is quite general, since it was shown by Rehfeld, that the differences between the isotherms shown by SDS at alkane/water interfaces are not very large, for liquid hydrocarbons between 10 and 17 carbons. Moreover, the method allows obtaining useful relationships to approximate the surfactant distribution of an ionic surfactant between the aqueous phase and the interface:

$$C_{eq}^{bulk} = 0.994687 \; C_s - 8.3925 \; x \, 10^{-6} \quad (r^2 = 0.999998) \tag{9}$$

$$C_{eq}^{surface} = -2.61330 \; x \, 10^{+4} \left(C_s\right)^4 + 5.72825 \; x \, 10^{+2} \left(C_s\right)^3 - 4.62531 \left(C_s\right)^2 + 1.95599 \; x \, 10^{-2} \left(C_s\right) + 1.49977 \; x \, 10^{-6}$$
$$(r^2 = 0.999193) \tag{10}$$

$$C_{eq}^{surface} = -1.17163 \; x \, 10^{-6} \; \gamma + 6.18341 \; x \, 10^{-5}$$
$$(r^2 = 1.00000) \tag{11}$$

In the equations above, the surfactant concentration is expressed in moles/liter and the tension is in mN/m.

The value of $N_{s,i}$ resulting from the use of Eqs. (4)-(5) is different from the one obtained from $C_{eq}^{bulk}$ (Eq. (10)):

$$N_{s,i} = \frac{A_T}{A_T} \left(C_{eq}^{surface} \; V_T\right) = A_r \; \Gamma \tag{12}$$

For the particular emulsion considered in this work: $A_T = 2.181 \; x \, 10^{-10} \; m^2$ and $V_T = 1.48 \; x \, 10^{-11}$ liters. Hence, the surface excess ($\Gamma$) can be computed using equations (10) and (12) as a function of the total surfactant concentration.

Notice that eqs. (10) and (11) were deduced using Model 1 to reproduce the actual adsorption isotherm. Hence, their predictions regarding the interfacial properties of the drops are different from the ones of Model 1 alone (see Figure 1). Thus, equations (9) to (12) themselves constitute an alternative algorithm for the calculation of the surfactant population at the interface. Hence, we shall refer to these equations as Model 2.

The results of Aveyard et al, and the ones deduced above from a combination of the adsorption isotherm of Rehfeld and our algorithm of homogeneous surfactant distribution, correspond to macroscopic systems. Therefore, Model 2 only translates the information obtained from the isotherm of Rehfeld to the surface of emulsion drops, taking into account the *discrete* contribution of each surfactant molecule.

Recently, James-Smith et al. [James-Smith, 2007] employed a very elaborate experimental methodology to quantify the distribution of SDS in 1% hexadecane/water nanoemulsions (600 nm < $R_i$ < 200 nm). The surfactant concentration ranged from 8 to 200 mM. The technique relied on the use of ultrafiltration tubes with nanoporous filters. Using a centrifugal force of 900 g, filters with a cut-off molecular weight of 30,000 only allow the passing of the aqueous solution of monomers and micelles, leaving a plug of hexadecane drops on the other side. The SDS concentration in the filtrate was quantified by electrostatic complexation with methylene blue. Following this procedure a plot of $C_{eq}^{bulk}$ vs. $C_s$ was obtained. It showed a straight line similar to the one of Figure 3, but with a slope of 0.5999 ($r^2 = 0.9964$), indicating that 40% of the surfactant was absorbed to the interface of the drops, regardless of the total surfactant concentration. Unfortunately, these experimental evaluations correspond to surfactant concentrations above the CMC, but the reliability of the technique was previously tested using aqueous surfactant solutions between 1 and 50 mM SDS.

Notice that if the slope of $C_{eq}^{bulk}$ vs. $C_s$ is less than 1.00 (Figure 3), the amount of surfactant adsorbed in nanoemulsions should be higher than the one deduced from the data of macroscopic interfaces. This contrasts with the results obtained by De Aguiar et al. and James-Smith et al. However, some inconsistencies between the value of the total interfacial area ($A_T$) deduced from the surfactant surface excess and the one obtained from the mean droplet size resulting from light scattering measurements, were also observed. In any event, the results of James-Smith et al. suggest that there is a considerable difference between the amount of surfactant absorbed to the interface of the drops in nanoemulsions, and the one attached to macroscopic interfaces.

Moreover, using small angle neutron scattering data (SANS) on hexadecane in water nanoemulsions, Staples et al. [Staples, 2000] found that the SDS concentration at the interface of the drops increases linearly between 1 and 7 mM ($C_s$ = 1 - 7 mM), but then it levels off at around 3 x $10^{-10}$ moles/cm² at $C_s$ = 30 mM SDS. Similar results in the range 10 to 30 mM were observed by Oelhke et al. [Oehlke, 2008].

### 3.2 Distribution strategies for apportioning ionic surfactants to the surface of emulsion drops in the presence of salt

Despite the abundant data on adsorption isotherms found in the bibliography, it is unusual to find analytical forms as simple and accurate as the one of Eq. (8). In this respect, the group of Gurkov [Gurkov, 2005] obtained a generalization of this relationship for the interfacial tension of a hexadecane/water interface in the presence of SDS and NaCl:

$$\gamma = 0.0401 \left(\ln\left(a_s \, a_t\right)\right)^3 + 0.7174 \left(\ln\left(a_s \, a_t\right)\right)^2 - 6.9933 \left(\ln\left(a_s \, a_t\right)\right) - 89.1414 \quad \text{(mN/m)} \quad (13)$$

Here $a_s$ and $a_t$ stand for the activities of the surfactant and the total concentration of electrolyte:

$$a_s = \gamma_\pm C_s \quad (14)$$

$$a_t = \gamma_\pm (C_s + C_{NaCl}) \quad (15)$$

$C_{NaCl}$ = [NaCl] in moles/liter, and $\gamma_\pm$ is the mean coefficient of activity, as described by the theory of Debye-Hückel:

$$\gamma_\pm = \frac{A \sqrt{I}}{1 + Bd_i \sqrt{I}} + b \, I \quad (16)$$

Here, I stands for the ionic strength; for NaCl solutions at 25 C, A = 0.5115 M$^{-1/2}$, Bd$_i$ = 1.316 M$^{-1/2}$ (d$_i$ is the ion diameter of the respective couple of ions), and b = 0.055 M$^{-1}$. The value of the surface excess (and the surfactant interfacial area) resulting from the differentiation of Eq. (13) with respect to ln ($a_s a_t$) is:

$$\Gamma = \frac{0.001}{k_B \, T} \left(0.1203 \left(\ln\left(a_s \, a_t\right)\right)^2 + 1.4348 \left(\ln\left(a_s \, a_t\right)\right) - 6.9933\right)$$
$$\text{molecules} / \text{m}^2 \quad (17)$$

Where $A_s = \Gamma^{-1} (C_s)$, $A_{s,min} = \Gamma^{-1}$ (CMC). Unfortunately, the paper of Gurkov et al. neither report the values of the CMC at each salt concentration, nor does it specify the value of the interfacial tension at the CMC ($\gamma_c$). These two parameters are necessary in order complete an adsorption isotherm. As shown by Eq. (13) the presence of electrolytes alters the adsorption equilibrium considerably, lowering the CMC of the system, and increasing the maximum amount of surfactant adsorbed [Aveyard, 1987]. In this regard, Aveyard et al. demonstrated that in the case of a heptane/water interface, the Gibbs adsorption isotherm still describes the behavior of the SDS in the presence of a considerable amount of NaCl. Furthermore, they showed that in the absence of an alcohol, this system does not attain ultralow interfacial tensions, as observed in the case of Gemini surfactants like AOT.

Experience shows that the CMC of ionic surfactants at an oil/water interface is often very similar to the one at an air/water interface, due to the fact that this type of surfactants is insoluble in the oil phase. This is fortunate because the experimental data concerning air/liquid isotherms is considerably larger than the one of liquid/liquid systems, a fact probably related to the late development of accurate tensiometers for interfacial measurements.

Figure 4 shows a set of empirical correlations between the CMC and $C_{NaCl}$ for both surface and interfacial systems containing SDS [Corrin, 1947; Emerson, 1965; Oehlke, 2008; Palladino, 2011; Umlong, 2007, Aveyard, 1987]. Some of them are well known, and some others are our reasonable fits of the reported experimental data. We are particularly concerned with salt concentrations between 300 and 600 mM, usually required for the evaluation of the flocculation rate of an emulsion [Rahn-Chique, 2012]. However, the experimental data readily available rarely reaches such high concentrations. For the present purposes, we chose the relationship of Corrin and Harkins [Corrin, 1947] to carry on our calculations. This equation was originally formulated for air/liquid systems:

$$Log_{10}\, CMC \;=\; -0.45774\; Log_{10}\left(C_{NaCl}\right) - 3.2485 \quad (18)$$

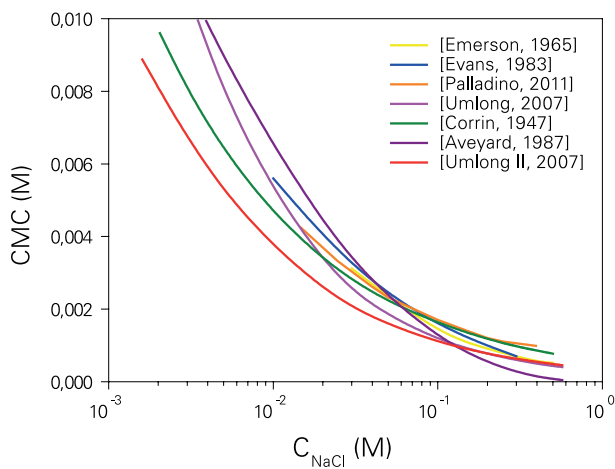

**Figure 4**: Behavior of several empirical equations for the prediction of the CMC in fluid/water systems as a function of the ionic strength of the aqueous solution.

In order to approximate $\gamma_c$, the relationship reported by Aveyard et al. [Aveyard, 1987] was used:

$$\gamma_c \;=\; -0.77969\; \ln\left(C_s + C_{NaCl}\right) + 3.1458 \quad (19)$$

In principle the use of equations (13) – (19) allow the calculation of the adsorption isotherms of SDS for any pair ($C_s$, $C_{NaCl}$). However, equations (13) and (17) are only valid over a restricted concentration range. Otherwise they produce negative values of either the tension or the average interfacial area. Since ESS must be able to provide an approximate surface potential for any arbitrary combination of $C_s$ and $C_{NaCl}$, two auxiliary equations were introduced in

the algorithm to correct this anomalous behavior **whenever it occurs**. For $C_s \geq 10^{-4}$ M:

$$\gamma \;=\; \gamma_0 + \left[\frac{\gamma_0 - \gamma_c}{\ln\left(C_{s,\min}\right) - \ln\left(CMC\right)}\right]\left(\ln\left(C_s\right) - \ln\left(C_{s,\min}\right)\right) \quad (20)$$

For $C_s < 10^{-4}$ M the values of $\gamma_c$ and CMC were substituted by $\gamma(C_s = 10^{-4}\,M)$, and $10^{-4}$ M, respectively.

In regard to the interfacial area, and independently of the surfactant concentration, $A_s$ was calculated from:

$$A_s = A_{s,\max} + \left[\frac{A_{s,\max} - A_{s,\min}}{\ln\left(C_{s,\min}\right) - \ln\left(CMC\right)}\right]\left(\ln\left(C_s\right) - \ln\left(C_{s,\min}\right)\right) \quad (21)$$

Where $A_{s,\max} = 100 * A_{s,\min}$.

The algorithm corresponding to the realization of Eqs. (13) to (21) is the generalization of Model 2 for the case in which there exists a finite salt concentration in the system. The isotherms resulting from this procedure are illustrated in Figure 5.

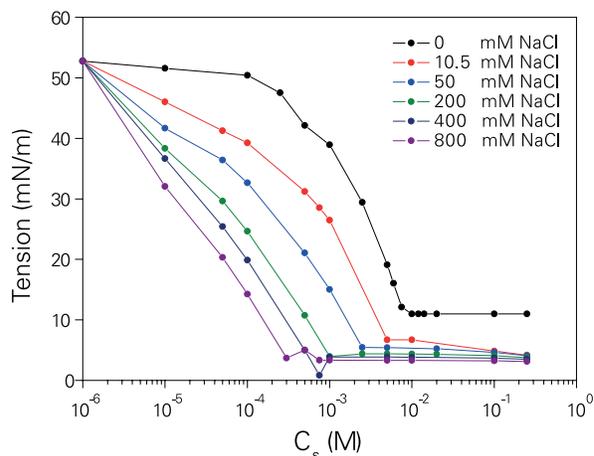

**Figure 5**: Generic isotherms predicted by Model 3 for the adsorption of SDS at an oil/water interface.

The approximate nature of the isotherms is clear from Figure 5. In particular, the anomalous minimum of the $C_{NaCl} = 400$ mM curve, and the lack of straight horizontal line above some CMC values, indicate that the subroutine can still be considerably improved. Yet, the values of the tension are expected to be close enough to those experi-mentally observed –at least for the purpose of ESS calculations-. These small abnormalities are related to Eqs. (18) to (21), that is, to our particular way of approximating the tension and the interfacial area when Eq. (13) and/or Eq. (17) fail to do so appropriately.

Once the "experimental" isotherms can be reasonably calculated for any pair ($C_s$, $C_{NaCl}$), we can repeat the procedure employed in section 3.1, and use Model 1 to reproduce the isotherms of Figure 5. We followed this laborious procedure in order to obtain the dependence of $C_{eq}^{surface}$ vs. $C_s$ at several salt concentrations. Figure 6 shows the data obtained for the case of $C_{NaCl}$ = 800 mM, along with its empirical fitting. Again, the variation of the surface concentration resembles a Langmuir isotherm (Figure 2), but in this case, it cannot be fit with a simple polynomial. Instead, it now approaches its saturation limit following a logarithmic trend:

$$C_{eq}^{surface} = A \ln \left( C_s \right) + B \qquad (22)$$

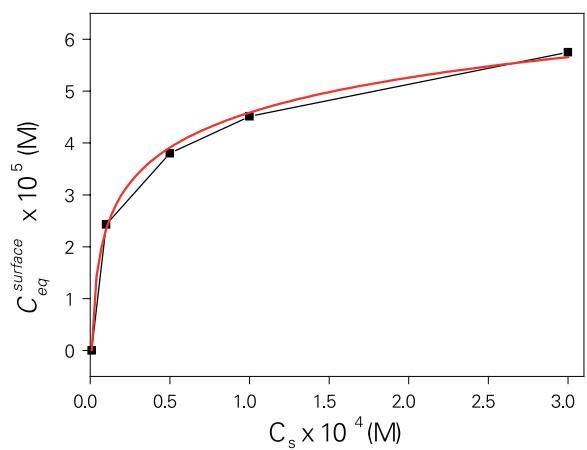

Figure 6: Surface concentration predicted by Model 3 for $C_{NaCl}$ = 800 mm.

Table 1 shows the regression coefficients of Eq. (22) for a selected set of salt concentrations.

It was also found that coefficients A and B can be conveniently expressed as second order polynomials on the salt concentration with reasonable regression coefficients:

$$A = -7.9700 \ x\,10^{-6} \ C_{NaCl}^{2} + 1.1496 \ x\,10^{-5} \ C_{NaCl} + \\ + 5.6724 \ x\,10^{-6} \qquad (r^2 = 0.96307) \qquad (23)$$

Table 1: Dependence of the coefficients of the Eq. (22) as a function of the NaCl concentrations.

| CNaCl (M) | A | B | r² |
|---|---|---|---|
| 0.05 | 6.21770 × 10⁻⁶ | 8.39520 × 10⁻⁵ | 0.97325 |
| 0.1 | 6.40590 × 10⁻⁶ | 8.79750 × 10⁻⁵ | 0.99083 |
| 0.2 | 8.15790 × 10⁻⁶ | 1.10730 × 10⁻⁴ | 0.98760 |
| 0.3 | 8.37050 × 10⁻⁶ | 1.14720 × 10⁻⁴ | 0.95240 |
| 0.4 | 9.15180 × 10⁻⁶ | 1.24750 × 10⁻⁴ | 0.99412 |
| 0.5 | 9.19610 × 10⁻⁶ | 1.26530 × 10⁻⁴ | 0.99585 |
| 0.6 | 9.51900 × 10⁻⁶ | 1.31150 × 10⁻⁴ | 0.99662 |
| 0.8 | 9.90110 × 10⁻⁶ | 1.37060 × 10⁻⁴ | 0.99813 |

$$B = -1.06704 \ x\,10^{-4} \ C_{NaCl}^{2} + 1.58345 \ x\,10^{-4} \ C_{NaCl} + \\ + 7.69257 \ x\,10^{-5} \qquad (r^2 = 0.97220) \qquad (24)$$

The set of equations (13) – (24) describe a new algorithm for the calculation of the interfacial surfactant population. In the following we shall refer to this algorithm as Model 3. Model 3 is expected to make reasonable predictions of the SDS surface concentration which are consistent with the experimental adsorption isotherm to a hexadecane/water interface in the presence of different concentrations of NaCl.

## 4. EXPERIMENTAL MEASUREMENTS AND COMPUTATIONAL PROCEDURE

### 4.1 Zeta potential measurements

In order to compare the surfactant surface excess predicted by the Gibbs adsorption isotherms (Model 3) with the one actually found in nanoemulsions, a set of oil-in-water dispersions was prepared for several salinities and surfactant concentrations. The oil employed was a mixture of 70% w/w hexadecane and 30% w/w tetrachloroethylene. This mixture is neutrally buoyant, and therefore suitable for ulterior turbidity measurements.

Previous to these measurements, the phase diagram of the system was obtained. This allowed identifying the regions of compositions where three-phases coexist. The mother emulsions were prepared by a sudden dilution of a three phase system [Solè 2006; García-Valero, 2014]. The same experimental technique was followed in all cases in order to obtain emulsions with similar Drop Size Distributions (DSD). The resulting DSD were measured using light scattering (LS 230, Beckman-Coulter) in order to decide if they were suitable for mobility measurements. The emulsions were discarded if either the average diameter of the drops fell outside the 350 - 380 nm range, or the polydispersity surpassed 20%. The final salt and the surfactant concentration were adjusted during the dilution of the mother emulsion. One emulsion was prepared for each ($C_s$, $C_{NaCl}$) pair. The number of particles in the dilution was approximately $3.6 \times 10^{10}$ particles/cm³. A minimum of three mobility measurements was done for each physicochemical condition using a Delsa 440 SX from Beckman-Coulter Co. The software of the equipment directly provides an average value for the zeta potential of the emulsion drops up to $C_{NaCl} \sim 0.2$ M. Above this concentration the temperature of the sample cell raises steeply and bubbles start to appear.

### 4.2 Adjustment of the surfactant charge ($q_s$) in order to reproduce the zeta potential of hexadecane/water emulsions using Model 1

In these calculations the total charge of a drop is supposed to be the result of the surfactant adsorption only (Eq. (7)). Thus, the surface charge density of drop i, $\sigma_i$ is simply:

$$\sigma_i = Q_i / \left(4 \pi R_i^2\right) \qquad (25)$$

The electrostatic surface potential of a drop of radius $R_i$, results from the numerical solution of Eq. (26) [Sader, 1997]:

$$\sigma_i \, e / \kappa \, \varepsilon \, \varepsilon_0 \, k_B T \;=\; \Phi_P + \Phi_P \left/\kappa \, R_i \right. - $$
$$- \kappa \, R_i \left(2 \sinh\left(\Phi_P/2\right) - \Phi_P\right)^2 \left/ \overline{Q} \right.$$
$$\overline{Q} \;=\; 4 \tanh\left(\Phi_P/4\right) - \Phi_P - \kappa \, R_i \left[2 \sinh\left(\Phi_P/4\right) - \Phi_P\right] \qquad (26)$$

Where: $\Phi_P = \Psi_0 \, e \, / \, k_B T$ is the reduced electrostatic potential of the particle at its surface, $\varepsilon_0$ is the permittivity of vacuum and $\varepsilon$ the dielectric constant of water. For the present purposes, the surface potential $\Psi_0$ is supposed to be equal to the zeta potential obtained from electrophoresis.

Notice that the free energy of electrostatic interaction between two drops can be calculated if their values of $\Psi_0$ are known [Danov, 1993]:

$$V_E = \left(64 \pi \, C_{el} \, k_B T/\kappa\right) \tanh\left(e \, \Psi_{0i}/4 \, k_B T\right) \tanh\left(e \, \Psi_{0j}/4 \, k_B T\right) \mathrm{x}$$
$$\exp\left(-\kappa \, h\right) \left[2 \, R_i \, R_j /\kappa \left(R_i + R_j\right)\right] \qquad (27)$$

Where: $C_{el}$ the concentration of electrolyte, $\kappa^{-1}$ is the Debye length, and $e$ the unit of electrostatic charge. The DLVO potential results from adding the attractive potential to $V_E$. It is usually calculated employing Hamaker's expression [Hamaker, 1937]:

$$V_A = V_{vdW} = -A_H/12 \left(y/\left(x^2 + xy + x\right) + y/\left(x^2 + xy + x + y\right)\right.$$
$$\left. + 2 \ln\left[\left(x^2 + xy + x\right)/\left(x^2 + xy + x + y\right)\right]\right) \qquad (28)$$

Here: $x = h/2 \, R_i$, $y = R_i/R_j$, $h = R_{ij} - R_i - R_j$, and $A_H$ is the Hamaker constant ($\sim 5.40 \times 10^{-21}$ J for pure hexadecane).

Although it is not the main purpose of this report, we will use Eq. (27) – Eq. (28) to illustrate how the differences in the surfactant surface excess predicted affect the interaction potential between the drops (and consequently the ESS predictions regarding the stability of the system studied).

## 5. RESULTS AND DISCUSSION

### 5.1 Predictions of the zeta potential based on Model 3

Models 2 and 3 are a reasonable ESS-implementation of the Gibbs adsorption equation for a typical alkane/water interface stabilized with SDS. However, we showed in section 3 that there is a controversy regarding the actual value of the surfactant surface excess in nanoemulsions. On the one hand the results of De Aguiar [De Aguiar, 2011] suggest that the surfactant population is much lower than the one exhibited at macroscopic interfaces. On the other the results of James-Smith suggest a much larger surface excess. Models 2 and 3 were deduced to reproduce the experimental data of macroscopic hexadecane/water interfaces. The question we wish to address is: How the electrostatic surface potential predicted by these models compare to the average value of the zeta-potential obtained from hexadecane-in-water nanoemulsions stabilized with SDS at different salt concentrations?

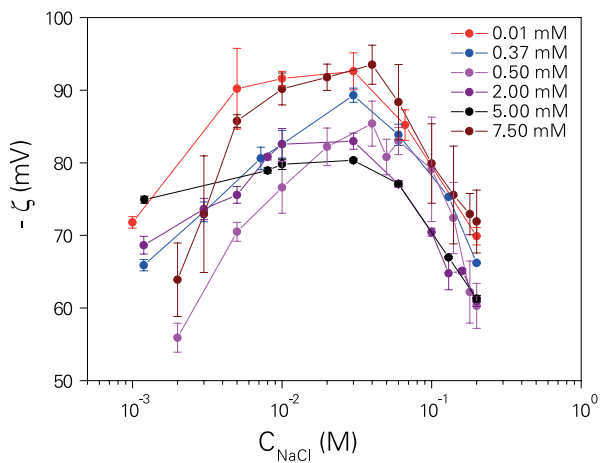

Figure 7: Zeta potential of hexadecane-in-water nanoemulsion drops as a function of the ionic strength at several surfactant concentrations.

Figure 7 illustrates the variation of the zeta potential of hexadecane-in-water nanoemulsions as a function of the salt concentration. Each curve in Figure 7 corresponds to a fixed surfactant concentration. In all cases it is observed that for low salt concentrations the zeta potential increases, passes through a maximum, and then decreases. This trend is reproducible despite the fact that the errors bars of the potential are very large. The observed behavior is expected on the basis of Debye Hückel theory. At low ionic strength the surface excess increases because the surfactant is "salted out" (the affinity of its hydrocarbon chain for the aqueous solution decreases with the increase of the salt concentration). This promotes the adsorption of the molecule to the interface. However, as the ionic strength of the solution increases the surface charge of the adsorbed surfactant molecules is screened, and the electrostatic surface potential decreases.

Notice that some of the curves appear to converge at high salt concentration. This is consistent with the fact that a maximum salt concentration implies both a maximum surface excess and a maximum screening of this surface charge by their counter ions.

A closer look at Figure 7 indicates that there appears to be no systematic order of the curves in regard to the surfactant concentrations employed. However, the emulsions corresponding to 0.5, 7.5 and 9 mM SDS were synthesized and prepared by a different operator one year before the last set of measurements was made, using a distinct stock of chemicals and a distinct calibration of the instrument. Whether this fact affects the order of the curves is unknown, but it is clear that these three curves break the monotonous trend shown by the rest of the curves. The rest of the plots suggest that the zeta potential of the emulsions decrease with $C_s$. This is consistent with the fact that the ionic strength of the aqueous solution increases with the total surfactant concentration employed, favoring the screening of the surface charge of the drops at high surfactant concentrations.

In our previous theoretical works we chose specific values of the surfactant concentration and the ionic strength in order to parameterize the surfactant charge. The charge $q_s$ (Eq. 7) was varied until the experimental value the zeta-potential ($\Psi i$) of the emulsion studied was reproduced for a selected pair of $C_s$ and $C_{NaCl}$ concentrations (Eqs. 25 – 26). Following this procedure, a value of $q_s$ = -0.09 was found to be necessary in order to reproduce the zeta potential ($\xi$ = -78.6 mV) corresponding to a drop of a hexadecane in water nanoemulsion with 7.5 mM SDS at an ionic strength (I) of 10 mM. This value of $q_s$ is somewhat low in comparison to our previous parameterizations of the charge (~ 0.21 [Urbina-Villalba, 2006]). In any event, the use of Model 3 along with a fixed surfactant charge of $q_s$ = 0.09 units, provides a first approximation to the experimental data of Figure 7 (see Figure 8).

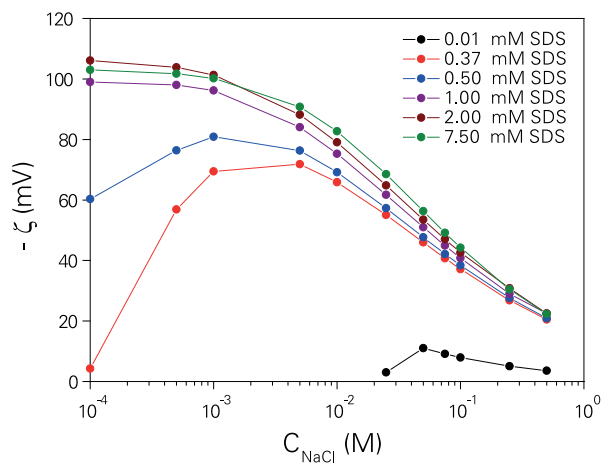

Figure 8: Model 3 predictions of the zeta potential of hexadecane-in-water nanoemulsion drops as a function of the ionic strength at several surfactant concentrations.

Notice that the zeta potentials exhibited in Figure 8 only show bell-shaped curves for very low surfactant concentrations. However, for $C_s$ > 1 mM SDS, the value of the potential decreases monotonically as the salt concentration increases. This trend does not change when the value of $q_s$ is varied between 0.06 and 0.30 (Figure 9). In general, the charge can be adjusted to reproduce the zeta-potentials corresponding to the higher salt concentrations, but notorious deviations are observed at lower ionic strengths. Moreover, the use of Model 1 or other routines for surfactant distribution available in our ESS code did not improve these results. In practical terms, the surfactant charge has to be high in order to produce high values of the zeta potential at high ionic strengths, but as a consequence, a minor adsorption at low salt concentrations, also generates high values of the zeta potential. The bell-shape curve is only recovered for very low surfactant concentrations, because the amount of surfactant adsorbed at low ionic strengths is also very low.

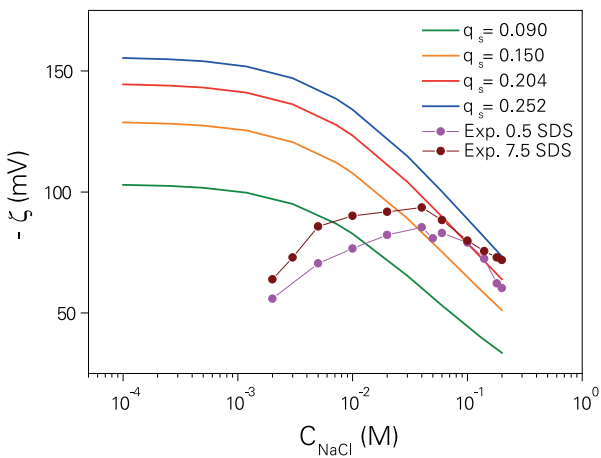

Figure 9: Model 3 predictions of the zeta potential of hexadecane-in-water nanoemulsion drops as a function of the ionic strength for several surfactant charges.

Hence, if we trust the reliability of Model 3 and the experimental values of the zeta-potential it predicts, the results of Figure 8 can only lead to one possible conclusion: the amount of surfactant molecules adsorbed at the interface of hexadecane drops of nanometer size ($R_i = 184$ nm) is significantly different from the one exhibited by macroscopic interfaces of the system under the same physicochemical conditions.

In order to reproduce the surfactant population suggested by the electro-kinetic measurements, we used Model 1 and followed the same methodology previously employed to obtain Models 2 and 3. In the former cases the adsorption isotherms (curves of $\gamma$ vs. $C_s$) were replicated at several ionic strengths. Here, we attempted to recover the variation of the zeta potential of the drops as a function of the ionic strength of the solution ($\zeta$ vs. $C_{NaCl}$) at several surfactant concentrations. For this purpose the value of $q_s$ adequate for reproducing the maximum value of the zeta potential at $C_s = 0.5$ mM SDS was selected ($q_s = 0.26$).

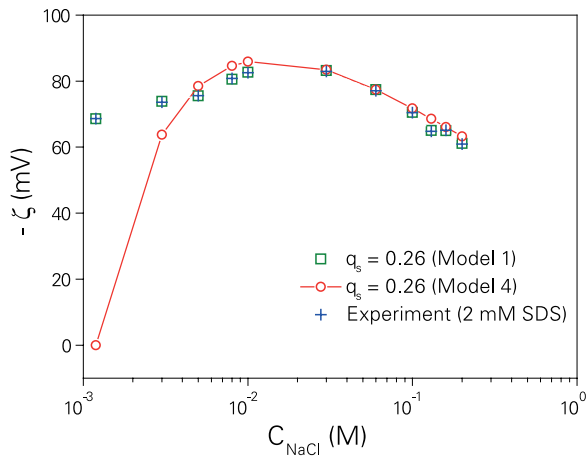

Figure 10: Model 4 predictions of the zeta potential of hexadecane-in-water nanoemulsion drops as a function of the ionic strength for $C_s = 2$ mM.

Figure 10 shows that in most cases, it is possible to use Model 1 in order to reproduce the experimental data exactly if the total surfactant concentration is arbitrarily changed. Small significant differences were only observed for the case of $C_s = 7.5$ mM at high ionic strength.

As expected, and despite the form of the curves of $\zeta$ vs. $C_{NaCl}$, the surface concentration obtained from these data monotonically increased as a function of the ionic strength. Table 2 shows the results of fitting the surface concentration with an expression of a type:

$$C_{eq}^{surface} = D \ln C_{NaCl} + E \qquad (29)$$

The regression coefficients are rather poor, and the dependences of D and E on the surfactant concentration are not monotonous, but instead correspond to very flat parabolas:

$$D = 0.1807868 \, C_s^2 - 1.464861 \times 10^{-3} C_s + 1.153241 \times 10^{-5}$$
$$(r^2 = 0.9848232) \qquad (30)$$

$$E = 0.9203015 \, C_s^2 - 7.052329 \times 10^{-3} C_s + 7.231916 \times 10^{-5}$$
$$(r^2 = 0.9975824) \qquad (31)$$

Table 2: Dependence of the coefficients of the Eq. (29) as a function of the surfactant concentration.

| Cs (mM) | D | E | r² |
|---------|---|---|-----|
| 0.5 | $1.09309 \times 10^{-6}$ | $6.91989 \times 10^{-5}$ | 0.96633 |
| 2.0 | $9.16225 \times 10^{-6}$ | $6.15600 \times 10^{-5}$ | 0.98281 |
| 5.0 | $8.84774 \times 10^{-6}$ | $6.03112 \times 10^{-5}$ | 0.96743 |
| 7.5 | $1.06732 \times 10^{-6}$ | $7.11073 \times 10^{-5}$ | 0.96426 |

Hence, it was preferred to compute an average curve independent of $C_s$. For this purpose an average value was computed averaging the predictions of Eq. (29) with the coefficients of Table 2 for each salt concentration. Curiously, averaging the coefficients of each fit naively (D = 9.903504 x 10⁻⁶, and E = 6.554437 x 10⁻⁵) fits the average data perfectly (Figure 11).

Equation (29) with constant coefficients (D = 9.903504 x 10⁻⁶, E = 6.554437 x 10⁻⁵) will be referred in the following as Model 4. Figures 10, 12, 13 and 14 show the predictions of the ζ-potentials produced by Model 4. The qualitative shape of the experimental curves is now reproduced. Comparison of the values predicted by Models 3 and 4 for $C_{eq}^{surface}$ illustrates a complex scenario: a) For $C_s$ = 2 mM or 5 mM, the surface concentration predicted by Model 4 is always higher than the one of Model 3, b) For $C_s$ = 0.5 mM SDS, the predictions of Model 3 are lower than the ones of Model 4 below $C_{NaCl}$ = 50 mM, and higher otherwise, c) For $C_s$ = 7.5 mM SDS, the predictions of Model 3 are lower than the ones of Model 4 below $C_{NaCl}$ = 50 mM, but similar otherwise.

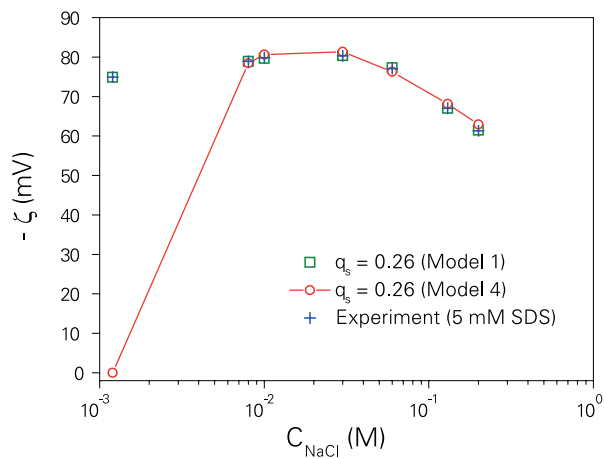

Figure 12: Model 4 predictions of the zeta potential of hexadecane-in-water nanoemulsion drops as a function of the ionic strength for $C_s$ = 5 mM.

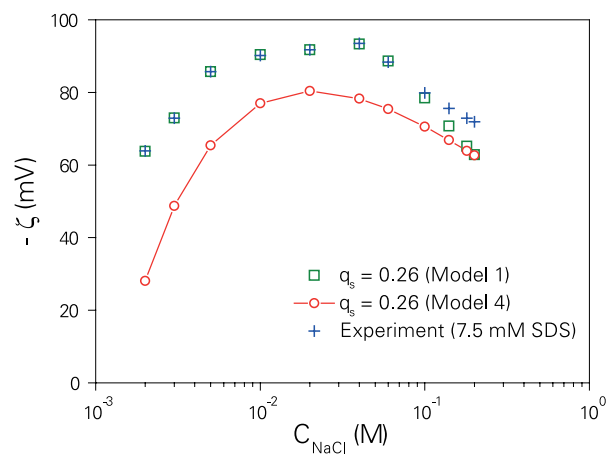

Figure 13: Model 4 predictions of the zeta potential of hexadecane-in-water nanoemulsion drops as a function of the ionic strength for $C_s$ = 7.5 mM.

## 5.2 Predictions of Models 3 and 4 for the free energy of interaction between two emulsion drops

Model 4 allows calculating a surfactant surface excess which is consistent with the experimental data of the zeta potential of hexadecane/water nanoemulsions stabilized with SDS, underline{regardless of the origin of the adsorption phenomenon involved}. Instead, Model 3 is our implementation of the Gibbs adsorption isotherm for the prediction of the SDS surface excess in these emulsions. The predictions are *very* different.

In order to evaluate if the differences in the predictions of Models 3 and 4 for the SDS surface excess may have consequences in the simulation of the stability of these emulsions, the interaction potentials between two drops of $R_1$ = 184 nm was calculated using equations (27) and (28). These potentials are shown in the Figures 15 through 18 for $C_s$ = 0.5 and 7.5 mM.

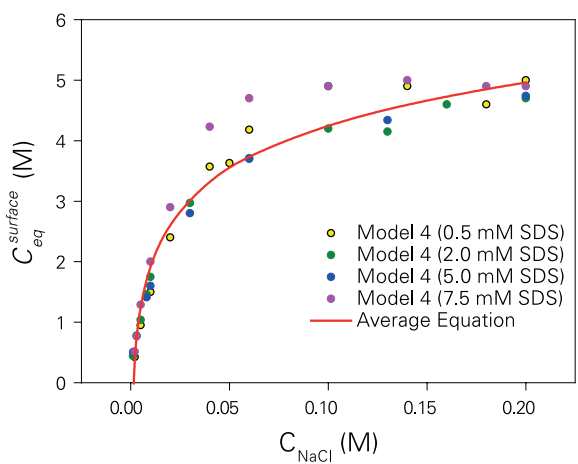

Figure 11: Predictions of Eq. (29) For the variation of the surface concentration hexadecane-in-water nanoemulsion drops as a function of the salt concentration of the system.

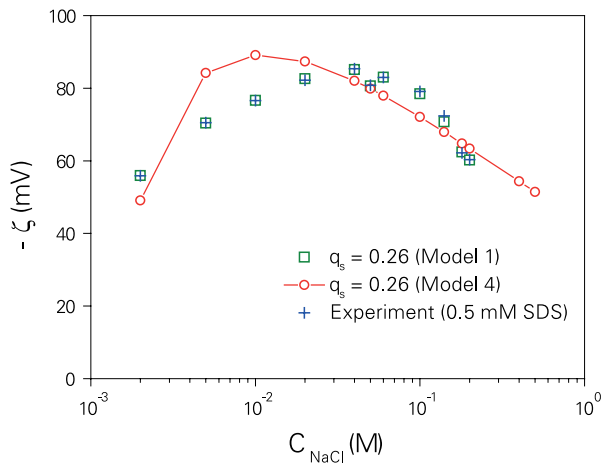

Figure 14: Model 4 predictions of the zeta potential of hexa-decane-in-water nanoemulsion drops as a function of the ionic strength for $C_s$ = 0.5 mM.

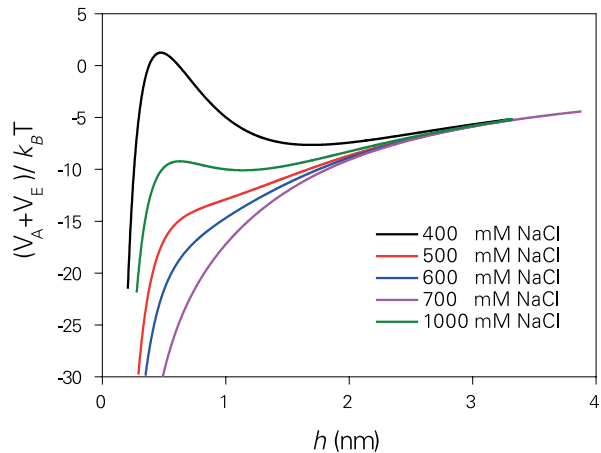

Figure 16: DLVO interaction energy predicted by Model 3 for two nanoemulsion drops of hexadecane suspended in water at a surfactant concentration of 7.5 mM.

Following the classical DLVO analysis, Model 3 (Figure 15-16) suggests, that the prepared nanoemulsions become unstable around $C_{NaCl}$ = 400 mM (= Critical Coagulation Concentration). Above this salt concentration the drops of the emulsions flocculate and coalesce (if they do not deform during the process of aggregation). Instead, Model 4 suggests (Figures 17-18) that the repulsive barriers are too high in all cases, and hence, only secondary minimum flocculation is possible at these ionic strengths. These two predictions are contradictory, and therefore, it appears feasible to test their reliability comparing ESS results with the actual stability of these systems.

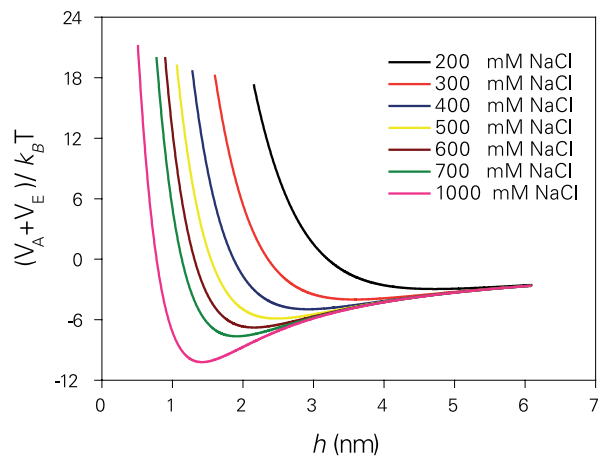

Figure 17: DLVO interaction energy predicted by Model 4 for two nanoemulsion drops of hexadecane suspended in water at a surfactant concentration of 0.5 mM.

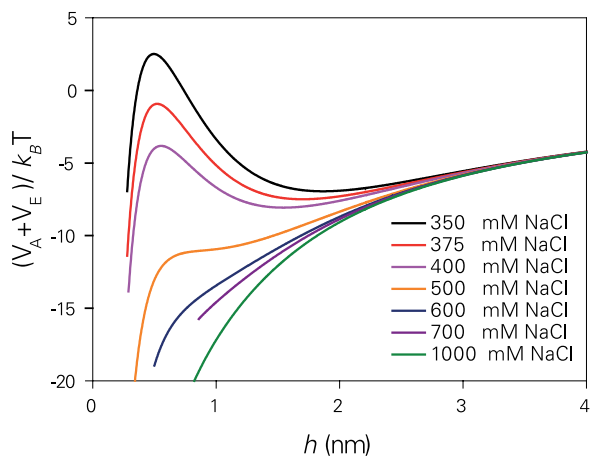

Figure 15: DLVO interaction energy predicted by Model 3 for two nanoemulsion drops of hexadecane suspended in water at a surfactant concentration of 0.5 mM.

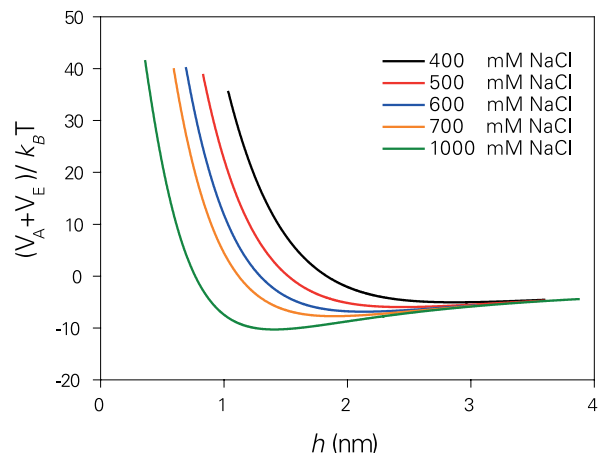

Figure 18: DLVO interaction energy predicted by Model 4 for two nanoemulsion drops of hexadecane suspended in water at a surfactant concentration of 7.5 mM.

## 6. CONCLUSION

A simple procedure for the calculation of the surfactant surface excess in oil/water emulsions was developed. Our results suggest that the surfactant surface concentration of hexadecane-in-water nanoemulsions stabilized with SDS is significantly different from the one predicted by experimental adsorption isotherms. Moreover, it can be higher or lower than the ones of macroscopic systems depending on the total surfactant concentration and the ionic strength.

## 7. ACKNOWLEDGEMENTS